\documentclass[runningheads]{llncs}

\usepackage{cite}
\usepackage{amsmath,amssymb,amsfonts}
\usepackage{graphicx}
\usepackage{textcomp}
\usepackage{url}
\usepackage{xcolor}
\usepackage{multirow}

\usepackage{algorithmic}%
\usepackage[ruled,linesnumbered,boxed]{algorithm2e}
\usepackage{graphicx}
\usepackage{subcaption}
\usepackage[english]{babel}
\usepackage{hyperref}
\usepackage{ctable}
\def\BibTeX{{\rm B\kern-.05em{\sc i\kern-.025em b}\kern-.08em
    T\kern-.1667em\lower.7ex\hbox{E}\kern-.125emX}}

\begin{document}
\title{MONDEO: Multistage Botnet Detection}

\titlerunning{MONDEO}

\author{Duarte Dias\inst{1} \and Bruno Sousa\inst{1} \and Nuno Antunes\inst{1}}

%
\authorrunning{D. Dias, B. Sousa and N. Antunes}


\institute{University of Coimbra, CISUC, DEI \ \\
\email{duartedias@student.dei.uc.pt, bmsousa@dei.uc.pt,nmsa@dei.uc.pt}
}

\maketitle              
\begin{abstract}
Mobile devices have widespread to become the most used piece of technology. Due to their characteristics, they have become major targets for botnet-related malware. FluBot is one example of botnet malware that infects mobile devices. In particular, FluBot is a DNS-based botnet that uses Domain Generation Algorithms (DGA) to establish communication with the Command and Control Server (C2).
MONDEO is a multistage mechanism with a flexible design to detect DNS-based botnet malware. MONDEO is lightweight and can be deployed without requiring the deployment of software, agents, or configuration in mobile devices, allowing easy integration in core networks.
MONDEO comprises four detection stages: Blacklisting/Whitelisting, Query rate analysis, DGA analysis, and Machine learning evaluation. It was created with the goal of processing streams of packets to identify attacks with high efficiency, in the distinct phases. MONDEO was tested against several datasets to measure its efficiency and performance, being able to achieve high performance with RandomForest classifiers.
The implementation is available at \href{https://github.com/TLDart/mondeo}{github}.

\end{abstract}
\section{Introduction}
\label{sec:Intro}

Mobile technology has been a massive success since its introduction, amassing a large portion of the active devices on the internet. These devices are low-powered, portable, and can use both LAN and WAN, giving them great versatility.

The security aspect of technology has been historically neglected. Mobile devices are no different. Even with the introduction of measures such as Google Play Protect\cite{GooglePlay} and Apple App Security~\cite{AppleAppSec}, malware is still able to sneak its way into users' devices. 

Mobile-related malware can occur under many formats and with many objectives, data exfiltration and financial damage, as mentioned in \cite{QAMAR2019887}. In particular, this document focuses on DNS-based malware that uses Domain Name Generation (DGA) algorithms as an evasive tactic, to hide its communications with Command and Control (C2) server(s). Diverse solutions for DGA approaches are available, such as Intel DGA~\cite{DGAIntel} and DGA Detective~\cite{DGACOSSAS}, having different performance levels and requirements. 

\subsection{FluBot}
In 2021, large amounts of botnet-focused malware started spreading on the internet. This malware can perform ransomware attacks and Distributed Denial of Service (DDoS), among other malware-specific functionalities~\cite{Prodaft21, Lyman_Interisle22}. \texttt{FluBot} was one of the most reported, mainly due to its massive lateral spreading capabilities. 

\texttt{FluBot}-infected applications act receivers for a Command and Control (\texttt{C2}) network. Upon installation, the malware first hides from the user using several evasive tactics, in order to ensure its long stay in the infected device. Its removal is also non-trivial, requiring a full device wipe. According to ProDaft's report \cite{Prodaft21}, \texttt{FluBot} can emulate well-known banking sites, steal 2-Factor Authentication (2FA) codes sent via SMS, exfiltrate data, uninstall applications, perform actions on users' behalf, among other vulnerabilities. 

Botnets, such as \texttt{FluBot} impact not only end-users but also network operators. Internet service providers (ISPs) having users with infected devices suffer from increased network traffic, reducing the overall quality of the network. For instance, Domain Generation Algorithms (DGA) perform large amounts of DNS queries before connecting to the Command and Control- C2 Server(s)~\cite{Lyman_Interisle22}.

\subsection{Malware Detection}
There are several methods to detect malware-infected systems. The static analysis comprises the methods that can be used without running the malware. Analyzing the malware's checksum is a form of static analysis. If the malware had been previously identified, then the hash comparison would quickly identify the presence of the identified malware. More in-depth static analysis dissects the program, analyzing its code, in order to understand its behavior.

The dynamic analysis consists of evaluating the malware while it is running. This practice is more dangerous and should only be performed in a safe environment (a sandbox). The dynamic analysis allows the researcher to identify runtime behaviour that may otherwise be impossible to find. To combat dynamic analysis, some malware has evolved to avoid running if it detects it is running in a virtualized environment.

Signature-based malware detection aims to identify the malware by relying on a list of well-known behaviours (these include, network traffic patterns, file hashes, system calls, etc). Signature-based detection usually provides accurate results at the expense of speed of detection (can only detect malware whose behaviour has been previously documented).

Anomaly-based detection expands the concept of signature detection, detecting malware based on unusual activity patterns. Such patterns include network traffic, connection timing, and system behaviour through the analysis of system calls. Anomaly-based detection does not need a previous baseline to detect new malware, granting better results on dynamic malware changes, at the cost of lower accuracy rates.

According to Singh et al., \cite{SINGH201928}, DNS-based botnet malware can be divided into five categories botnet: Anomaly, Flow, Flux, DGA, and Bot Infection. Following this classification, Flutbot is considered to be a DGA-based malware.

AI has proven to be a valuable method for the detection of botnets. Examples include HANABot~\cite{Almutairi2020}, MABDS~\cite{MADBDDS14} or BotMark~\cite{BotMark-Wang2020}, which make use of techniques such as Reinforcement Learning, Adaboost, K-Nearest Neighbors (kNN), Support Vector Machines (SVM), and Isolated Forests.

\subsection{MONDEO}

\texttt{MONDEO} was implemented as a proof-of-concept (POC), using a docker container running python-flask, using RESTful API responses, and was evaluated with datasets both created in the laboratory and by collecting data on a deployed DNS server. The results show the suitability of MONDEO to be deployed in network infrastructures of operators with minimal overhead and with accurate precision levels.

Regarding performance, \texttt{MONDEO} aims to be time efficient, balancing accuracy and speed of detection to deliver precise results using production-grade scenarios as a baseline for tool design.

\texttt{MONDEO} provides a solution that does not require software deployment, agents, or configuration in mobile devices. To achieve this, it comprises a flexible multistage pipeline, which processes streams of packets, providing a per-request floating-point classification (0 to 1, Non-infected to Infected). This approach combines anomaly detection, DGA-based detection, and machine learning in an effort to obtain better and faster results. MONDEO also innovates in the aspect that the results of some phases can be used to configure the initial phase, where blacklists and/or whitelists are applied.

The solution of MONDEO is available at \href{https://github.com/TLDart/mondeo}{github}.

\subsection{Structure}

The document considers the following structure. Section~\ref{sec:soa} presents an overview of the state-of-the-art botnet detection, as well as relevant literature on the Flubot malware. The section concludes by comparing the MONDEO approach to other state-of-the-art approaches.
Section~\ref{sec:mondeo} details the MONDEO pipeline, carefully describing each step; 
Section~\ref{sec:evalResults} presents the evaluation methodology, while Section~\ref{sec:evalResults:results} documents the achieved results; 
Section \ref{sec:fw} discusses possible changes and additions to be made to MONDEO; 
Section \ref{sec:conclusions} concludes the document providing an overview of the technology as well as a final analysis.

\section{Related Work and MONDEO positioning}
\label{sec:soa}
This section presents the related work and positions MONDEO regarding the SoA.

\subsection{Related Work}
Khraisat et al.~\cite{Ansam19} provide an overview of Network-based Intrusion Detection Systems (NIDS), Host-based Intrusion Detection systems (HIDS) Signature-based detection (SIDS), and Anomaly-Based detection (AIDS). It also provides an overview of algorithms and techniques used in machine learning detection approaches.

Manmeet Singh et al.~\cite{SINGH201928} survey DNS-based botnet detection frameworks, documenting the usage of different techniques. This work divides detection into five categories: anomaly-based, flow-based, flux-based (e.g., using IP information), DGA-based, and Bot infection detection-based. According to the research, the majority of the works identified rely on data for DNS response-based features, which take a longer time to perform accurate detection.

Majda Wazzan et al.~\cite{Wazzan2021} survey botnet detection mechanisms for IoT devices, including connected cameras, routers, and Android devices. For threats such as Mirai and Reaper, detection techniques are documented for different malware phases such as Reconnaissance, Spread, and Attack. MONDEO shares similarities in detection techniques used such as Adaboost, K-Nearest Neighbors (kNN), Support Vector Machines (SVM), and Isolated Forests. Noticeably, detection using such techniques is performed during the attack phase, where infected devices establish communication with the C2 server and act actively in attacks.

Ying Xing et al. \cite{Xing2021} survey botnet detection techniques including honeypot analysis, communication signatures (e.g. using white and blacklists), Deep Learning techniques (both based on Neural networks or based on Reinforcement Learning), statistical analysis, distributed approaches and also combination methods. The work includes recent approaches for Moving Target Defense (MTD), also detailing combination methods like HANABot~\cite{Almutairi2020} or MABDS~\cite{MADBDDS14} which use multiple techniques to detect botnet requiring the deployment of agents in the mobile devices and at the network side like honeypot agents.

Rosa et al. \cite{ROSA202150} provide an overview of machine learning techniques for malware detection in Industrial Autonomous Control Systems (IACS). For detection using Artificial intelligence (AI), the techniques with the best results mentioned are k-Nearest Neighbour (kNN), Support Vector Machines (SVM), and Isolated Forests.

Salsabila et al.\cite{salsabila2022flubot} analyze FluBot using a combination of static and dynamic analysis. This paper focuses on disassembling the malware and analyzing the underlying code, as well as documenting the runtime behaviour of the malware in the infected device.

BotMark~\cite{BotMark-Wang2020} is a botnet detection strategy based on several techniques and with multiple steps. Techniques include pre-processing, hybrid analysis, and bot detection. The approach was created to be protocol independent and therefore, the used features do not depend on the protocol used, for example, packets per second (PPS) and standard deviation of payload size. BotMark employs kNN techniques for clustering and identifying distinct communication flows. While the evaluation achieves high accuracy in datasets with Mirai, Ares, and Black-Energy, BotMark does not include whitelists or blacklists, meaning it can provide false negatives in the traffic patterns of legitimate applications like Redis or Zookeeper with similar characteristics of malware in botnets.

\subsection{MONDEO - Innovation and approach comparison}

\texttt{MONDEO} is positioned as an approach that combines several phases with accuracy, efficiency and integration purposes. The multi-staged approach allows network operators to perform botnet detection using existing whitelist/blacklisting approaches, without requiring the installation of any in mobile phones. 

When compared to other approaches that can detect FluBot-based attacks, Salsabila et al. \cite{salsabila2022flubot} focus on static and runtime behaviour analysis, whereas \texttt{MONDEO} focuses on network-based detection. Bellizi et al. \cite{9737464} analyze memory dumps using the JIT-MF framework to detect malware in the users' system, such as FluBot. \texttt{MONDEO} does not focus on the user device but on identifying malware from the perspective of the mobile carrier, presenting a solution that does not imply any contribution from the infected user. Chiscop et al. \cite{chiscop2022ai} identify FluBot based on its DGA behaviour, \texttt{MONDEO} combines DGA detection with other features, such that detection can be performed more efficiently.

Regarding other network-related botnet detection approaches, Singh et al. \cite{SINGH201928} present solutions where both DNS replies/answers, where \texttt{MONDEO} only require the usage of DNS requests for detection.

Other works also propose a multilayered approach to botnet detection. Almutiari et al. \cite{Almutairi2020} propose HANABot, a machine-learning-based evaluation algorithm that uses features from both the network and the host. Wang et al. \cite{MADBDDS14} propose BBDP, behaviour-based botnet detection in parallel. This approach uses 5 stages to detect botnets, consisting of traffic reduction, feature extraction, data partitioning, DNS detection, and TCP detection. Of the mentioned stages, \texttt{MONDEO} shares only the DNS detection stage. 

The main \texttt{MONDEO}'s novelty is related not to the underlying detection principles used, but to how the approach is taken and how resources are combined to ensure efficient and accurate detection.

\section{MONDEO Implementation Details}
\label{sec:mondeo}

\begin{figure}[!htbp]
  \centering
	\includegraphics[width=0.82\columnwidth]{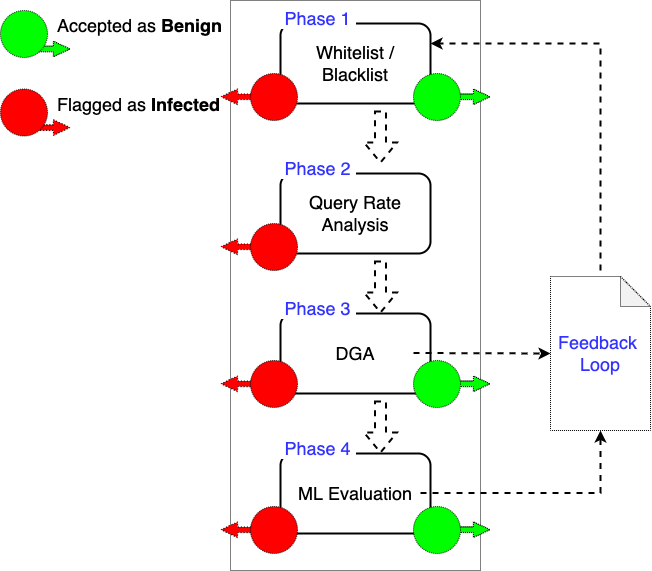}
	\caption{MONDEO overall stages and feedback loop}
	\label{fig:MONDEOoverall}
\end{figure}

MONDEO is a multistage pipeline configured for flexibility and efficiency. It contains 4 evaluation stages: 1- Whitelisting/Blacklisting; 2- Query Rate Analysis; 3- DGA Evaluation; and 4- Machine Learning Evaluation. The diverse stages rely on current practices regarding access control in networks like the whitelists/blacklists and requests ratios. The DGA and ML are specific to the Flubot malware behaviour. Figure \ref{fig:MONDEOoverall} illustrates the pipeline.

Each phase may generate an evaluation resulting in one of 3 actions: 
\begin{itemize}
    \item \textbf{Benign} when part of a regular request. This is present in all the phases, except phase 2 which does not do this classification.
    \item \textbf{Infected} when belonging to malware requests.
    \item \textbf{Pass for next stage} when a phase considers it does not have enough information, having the packet transit to the next phase.
\end{itemize}
The last phase with Machine Learning models always produces a definite evaluation, either Benign or Infected. 

The ML and DGA phases support a feedback loop which is able to automatically populate both white and blacklists, further improving efficiency, especially on the subsequent analysis, where requests can be evaluated in the first phase.

MONDEO uses solely DNS queries for botnet detection, which is relevant for mobile network operators, as most mobile users rely upon the default DNS servers on their devices, usually provided by the operator. Furthermore, by working on the side of the network operator, MONDEO removes the need for an end user to install agents, and applications to run on their devices.

\subsection{Phase 1 - Whitelisting/Blacklisting}
\label{sec:mondeo:stages:phase1}

A whitelist and blacklist correspond to lists where domains are catalogued and directly compared against the queried domain. If a domain is in the whitelist, the packet passes evaluation whereas if the packet is in the blacklist, the packet is immediately flagged as infected. For efficiency, a hash-based structure guarantees $O(1)$ search complexity. If the list is too big to be reliably implemented using hash-based data structures, binary structures are a good second option, with an associated $O(log(n)$ complexity for the search operation. 

The options described above take into consideration a full (1-1) direct match. To improve the performance, the Free Level Domain (FLD) can be stripped from the full query domain, and, while it has slightly lower precision, it doesn't affect results too much. For example, the domain
``\texttt{.netflix.com}'' contains several FLD's, namely \texttt{ftl.netflix.com}, \texttt{api-global.netflix.com}.

The Proof of Concept (PoC) elaborated to evaluate MONDEO employs a linear search with  $(O(n))$ complexity with FLD stripping. 

The whitelist and blacklist can be dynamically improved by adding domains to the list according to the runtime decision of other phases (such as phases 3 and 4) on those domains. That is through the feedback loop. 
Nonetheless, it should be noted that adding to a whitelist without manual confirmation can be dangerous from a security standpoint.

\subsection{Phase 2 - Query Rate Analysis}
\label{sec:mondeo:stages:phase2}

Analyzing the FluBot action pattern reveals that, to establish a connection with the C2 server, FluBot spams several thousands of queries in a short period of time until it connects with the real \texttt{C2} server. 

This pattern can be detected by implementing a mechanism to detect a high query rate. An efficient approach makes implementation nontrivial, as the best structure would require an event-driven data structure, such as a doubly-linked list, where both ends of the lists are updated whenever a new packet is processed, ensuring the time interval, $\Delta t$, is maintained.

A simpler but also efficient approach was used in the PoC. Instead of keeping track of all of the events (i.e. packets) at any given time window, the difference between any 2 contiguous packets is measured. For example, if \textit{Packet\_1} arrived at timestamp $t1 = 1$ and \textit{Packet\_2} arrived at $t2 = 2$ then they diverge in one time-unit, which cannot necessarily be measured in seconds. Using this approach sensitivity can be parameterized in 2 ways:
\begin{itemize}
    \item $\Delta F$ refers to the divergence interval in time units;
    \item $K$ specifies the threshold in the number of packets that surpass the divergence threshold.
\end{itemize}

If $\Delta F = 0$ then packets must arrive with the same timestamp (the smallest value possible and, therefore, the least of packets are caught). $K$ limits false positives delaying warnings. For example, in situations where a legitimate service makes 5 queries under $\Delta F$ time, it is not reported for any  $K \leq 5$. 
Algorithm~\ref{alg:QueryRate} summarises the steps that are performed in the Query Ratio analysis phase.

\begin{algorithm}[!tbp]
\small
\algsetup{linenosize=\tiny}
\SetAlgoLined
\KwIn{time of packets ($tpkt_{1}$ ... $tpkt_{n}$), $\Delta F$, $K$}
\KwOut{Decision if abnormal Request Ratio $flagInfected$}
\BlankLine
$\Delta T = tpkt_{n} - tpkt_{n-1}$ \\
$cnt = n$ \\
\eIf{$\Delta T \leq \Delta F$}{
  \eIf{$cnt \leq K$}{ 
  $flagInfected=true$}
  {$flagInfected=false$}
 } {$flagInfected=false$}
\caption{Query Rate and sensitivity configuration }
\label{alg:QueryRate}
\end{algorithm}

\subsection{Phase 3 - DGA Detection}
\label{sec:mondeo:stages:phase3}
Any packet passing phase 2 has its Fully Qualified Domain Name (FQDN) analyzed by a DGA checker. DGAs produce random-but-deterministic FQDNs names that are used in FluBot's DNS requests.
The attacker registers FQDN(s) as legitimate C2 server(s), and if that one gets blacklisted, the attacker quickly registers a new one. This technique is much more resilient than using static IPs.

In the MONDEO PoC we have considered two main implementations of the DGA detector: 
\begin{itemize}
    \item DGA Intel (\textit{DGAIntel}) ~\cite{DGAIntel}, maintained by Intel
    \item DGA Detective (\textit{DGADet})~\cite{DGACOSSAS}, result from the H2020 SOCCRATES research project.
\end{itemize}
The open-source solutions perform analysis of the DNS requests, resulting in a floating point evaluation where 0 indicates that a domain is non-DGA generated and 1 that it is DGA-Generated. The acceptance/rejection criteria were defined with lower and upper boundaries as follows:
\begin{itemize}
    \item  $lower \leq 0.1$ means immediate acceptance. 
    \item  $upper \geq 0.9$ delineates immediate rejection.
\end{itemize}

%
%
%
\subsection{Phase 4 - Machine Learning Detection}
\label{sec:mondeo:stages:phase4}
The last phase of the pipeline performs a machine learning evaluation. As expected this phase takes more time and should evaluate, fewer requests to reduce the impact in the detection of Flubot. 
This phase produces a binary output where 0 is not infected, and 1 is an infected packet.

\subsubsection{Feature Selection}
The features used are summarized in Table~\ref{table:features} and are taken from DNS requests. Some of the fields were converted to numeric values for efficiency concerns. The IP addresses used bit conversion for each decimal octet of an IP address version 4.
\ctable[
cap=Selected Feature Set,
caption={Selected Feature Set},
label=table:features,
pos=!htb,
doinside=\footnotesize,
center,
nostar
]{m{2.5cm} m{5cm} m{2.5cm}   }{
}{ \FL
\textbf{Feature ID} & \textbf{Description} & \textbf{ML Data Type}\ML 
IP Src & Source IP performing request  & Bit Conversion  \\
IP Dst & Destination of DNS request & Bit Conversion  \\
Length & Size of the Payload & Integer \\
DNS Flag & Info Regarding Flags & Boolean \\
DNS Questions & N. of requests in DNS message & Integer \\
Query Type & Qry Type: {\scriptsize A, AAAA, CNAME, PTR} & integer \\
Qry Name Null & If DNS name is NULL or not & Boolean \\
Timestamp & Indication packet creation time & Integer
 \LL
}

\subsubsection{Model Training}
\label{sec:MONDEOImpModelTrain}
The training data included a total of 10.000 data points, from which there is a 50/50 split between a fabricated packet set using the Alexa Top 1 million domain list~\cite{AlexaTop1M}, and lab-generated malware samples (see Section~\ref{sec:evalResults:DNSDataSets}). The Alexa Top 1 million was used since it contains the most well-known, or used domains in the Internet. The names were used to generate DNS requests with names of these domains, such as \texttt{www.facebook.com}.  

The trained ML model uses an 80/20 test/train split, where 80\% of the data is used to train the model and the remaining 20\% is used to test the accuracy of the model. 
The model is implemented in Python using the scikit-learn, using \textit{RandomForest},  \textit{IsolationForest}, \textit{MPLC}, and \textit{SVM} model classifiers. Such models provide an acceptable tradeoff regarding classification accuracy and performance~\cite{IsolationForestPerformance}.

\section{MONDEO Evaluation Methodology }
\label{sec:evalResults}
This section documents the evaluation methodology to assess the performance of MONDEO.%

\subsection{DNS Experimental Setup and Datasets}
\label{sec:evalResults:DNSDataSets}

To create a realistic dataset a DNS server was configured using ISC BIND. The information present in the datasets included captured DNS packets from the regular (non-infected) DNS clients.

The data collected comes from volunteers, which configured their devices to use the DNS server. About 20 users participated in the experiment with their mobile phones and laptops. The collection of DNS data was performed during a period of three months. 

To assess the behaviour of FluBot safely, Android Studio was used to sandbox the malware samples. The samples with FluBot malware were available in three distinct applications, UPS, Correos and DHL, as summarised 
in table~\ref{table:Malware}. 

\ctable[
cap=FluBot malware sample information,
caption={FluBot malware sample information},
label=table:Malware,
pos=!htb,
doinside=\footnotesize,
center,
nostar
]{m{2cm} m{2cm} m{7.0cm}   }{
}{ \FL
\textbf{Name} & \textbf{File(s)} & \textbf{Description}\ML 
UPS & 83 & Application that mimics official UPS app\\
Correos  & 108, Lab & Application that mimics Correos app \\
DHL  & 125 & Application similar to DHL app for tracking
 \LL
}

The profile for the emulated device was based on Google's Pixel 4, running the Android API 29. This device is also used in related works assessing the behaviour of mobile devices \cite{Rybakov_2020,Pixel4_9365835}. Malware was activated in specific time windows such that clear samples could be captured.

\subsection{Evaluation Methodology}
\label{sec:evalResults:evalMeth}

The MONDEO framework was evaluated under two parameters: the accuracy of detection and the efficiency of detection.

\subsubsection{Machine Learning Model}
\label{sec:mlModelAcc}

The goal is to verify the findings in the state of the art \cite{IsolationForestPerformance} regarding the applicability of the ML classifiers models like \textit{RandomForest},  \textit{IsolationForest}, \textit{MPLC}, and \textit{SVM}.

To assess the accuracy of the machine-learning models, we have used the metrics provided by \textit{scikit-learn} for the diverse models, as summarised in Table~\ref{table:PerfMetricsML}. 

\ctable[
cap=Metrics for ML Accuracy,
caption={Metrics for ML Accuracy},
label=table:PerfMetricsML,
pos=!htb,
doinside=\footnotesize,
center,
nostar
]{m{0.12\columnwidth} m{0.13\columnwidth} m{0.64\columnwidth} } {
}{ \FL
\textbf{Metric} & \textbf{Unit} &  \textbf{Description}\ML 
Precision  & $0 \leq x \leq 1$  & Ratio $TP / (TP + FP)$ where TP is the number of true positives and FP the false positives. 
\\
\hline
Recall & $0 \leq x \leq 1$ & Ratio $TP / (TP+FN)$ measuring the percentage of actual infected samples correctly classified. \\
\hline
F1-Score & $0 \leq x \leq 1$ & Determined by the formula $2* (\frac {Precision*Recall}{Precision+Recall} )$. 
1 represents the best score and 0 the worst.\\%
\hline
Accuracy & $0 \leq x \leq 1$ &  
Determined by matching predicted labels with \textit{y\_true} (real value). \\
\hline
Support & $ x \geq 0$ & Number of occurrences of each class in \textit{y\_true}.
\LL
}

\subsubsection{MONDEO Pipeline}
\label{sec:evalMeth:MONDEODataPipeline}
MONDEO's evaluation was performed in a virtual machine configured with 4vCPUS and 16GB of RAM. MONDEO's performance was evaluated individually in each phase in terms of processing time per phase and the overall number of packets processed. 

The solution is implemented in two dockers: 
\begin{itemize}
    \item \textbf{MONDEO core} assuring functionalities in Phase 1, Phase 2 and Phase 4.
    \item  \textbf{MONDEO DGA} which implements Phase 3 only, running the specific solutions of the DGA Intel and DGA Detective.
\end{itemize}
MONDEO core which in one containing the phases 

\ctable[
cap=Tests in MONDEO Pipeline Evaluation,
caption={Tests in MONDEO Pipeline Evaluation},
label=table:FilesEval,
pos=!htb,
doinside=\footnotesize,
center,
nostar
]{m{1cm} m{1.5cm} m{2.5cm} m{4.8cm}   }{
}{ \FL
\textbf{Test} & \textbf{Type} & \textbf{File(s)} & \textbf{Description}\ML 
\#1 & Infected & 83, 108, 125, Lab & With samples of malware  \\
\#2 & Benign & 23, 240, Alexa & Only with regular DNS requests 
 \LL
}

The test data used was retrieved from the DNS packet collection setup (summarised in Table~\ref{table:FilesEval}). In addition, test \#2 included a crafted sample, based on  \textit{Alexa Top 1 Million}~\cite{AlexaTop1M} list, which includes the most visited domains on the internet, as described in section~\ref{sec:mondeo}. The dataset of Alexa increases the number of the tested domains, adding reliability to the experiments.

The metrics used to assess the performance of MONDEO are summarised in Table~\ref{table:PerfMetricsDataPipeline}
\ctable[
cap=Performance Metrics in the MONDEO  Pipeline,
caption={Performance Metrics in the MONDEO  Pipeline},
label=table:PerfMetricsDataPipeline,
pos=!htb,
doinside=\footnotesize,
center,
nostar
]{m{2.8cm} m{1.5cm} m{6.8cm}   }{
}{ \FL
\textbf{Metric} & \textbf{Unit} &  \textbf{Description}\ML 
Packets Processed & \% & Ratio of packets processed in each phase, considering the total of captured packets\\
\hline
Processing Time & ms & Time to process a packet in each phase \\
\hline
Classification & n/a & Final classification of MONDEO, if the packet is flagged as Infected or is identified as Benign.
 \LL
}

To perform the measurement of time, the \textit{TimeIt} python package was included in the developed PoC. This module directly measures the time taken by Python methods, returning a 16-digit decimal value.

The resources consumed in terms of CPU and memory usage, as well as the amount of input and output traffic, are exchanged for the classification process. 
%

\section{Evaluation Results}
\label{sec:evalResults:results}

This section presents the evaluation results, according to the evaluation methodology, and chosen models.

\subsection{ML Model Accuracy}
\label{sec:evalResults:results:ML}

The ML model accuracy results are summarised in Table~\ref{table:MLAccuracyResults}. The model has higher accuracy and efficiency when the value of the diverse metrics is higher. The main goal of the distinction models is to assess the one that is more suited to MONDEO multistage pipeline.

\ctable[
cap=ML Accuracy of the Different Classifiers,
caption={ML Accuracy of the Different Classifiers},
label=table:MLAccuracyResults,
pos=!htb,
doinside=\footnotesize,
center,
nostar
]{m{2.2cm} m{2cm} m{1.9cm} m{1.4cm} m{1.4cm}  m{1.4cm} }{
}{ \FL
\textbf{Classifier} & \textbf{Category} & \textbf{Precision} &  \textbf{Recall} & \textbf{F1-score} & \textbf{Support}\ML 
\multirow{ 3}{*}{RandomForest} & Benign (0) & 1.00 & 1.00 & 1.00 & 1014 \\
 &  Infected (1) & 1.00 &  1.00 &  1.00 & 986 \\
 \cline{2-6}
 & Accuracy & & & 1.00 & 2000 \\
\hline
\multirow{ 3}{*}{SVM} & Benign (0) & 1.00 & 0.51 & 0.68 & 953 \\
 & Infected (1) & 0.69 &  1.00 &  0.82 & 1047 \\
  \cline{2-6} 
 & Accuracy & & & 0.77 & 2000 \\
\hline
\multirow{ 3}{*}{MLPC} & Benign (0) & 1.00 & 0.00 & 0.00 & 1021 \\
& Infected (1) & 0.49 &  1.00 &  0.82 & 979 \\
 \cline{2-6}
& Accuracy & & & 0.49 & 2000 \\
\hline
\multirow{ 3}{*}{IsolationForest} & Benign (0) & 1.00 & 0.00 & 0.00 & 955 \\
& Infected (1) & 0.57 &  0.75 &  0.65 & 1045 \\
 \cline{2-6}
& Accuracy & & & 0.39 & 2000 
\LL
}

In terms of accuracy, the \textit{RandomForest} is the classifier with the best performance, being able to correctly classify all the non-infected and infected samples. All the models are able to classify the non-infected samples but differ when classifying the infected samples, with MLPC providing the worst performance in terms of precision and accuracy. Given such results, the \textit{RandomForest} is chosen as the classifier to be used in the Pipeline. The results achieved with the ensemble modes like \textit{RandomForest} and  \textit{IsolationForest} are in line with the achieved results in SoA for anomaly detection~\cite{IsolationForestPerformance} (recall section~\ref{sec:MONDEOImpModelTrain}).

\subsection{MONDEO Pipeline Performance}
\label{sec:evalResults:results:Pipeline}
The results in the MONDEO Pipeline are presented considering tests summarised in Table~\ref{table:FilesEval}, for three approaches: 1) \textit{DGAInt} - DGAIntel without a feedback loop, 2) \textit{DGAInt-Fed} - DGAIntel with the feedback loop, 3) \textit{DGADet} - DGA Detective without a feedback loop.
It should be noted that the samples are not balanced, Test \#1 contains a majority of \textit{Infected} samples while Test \#2 contains only \textit{Non-infected} packets.

\begin{figure}[!hbt]
    \begin{subfigure}{0.5\textwidth}
    \centering
    \includegraphics[scale=0.4]{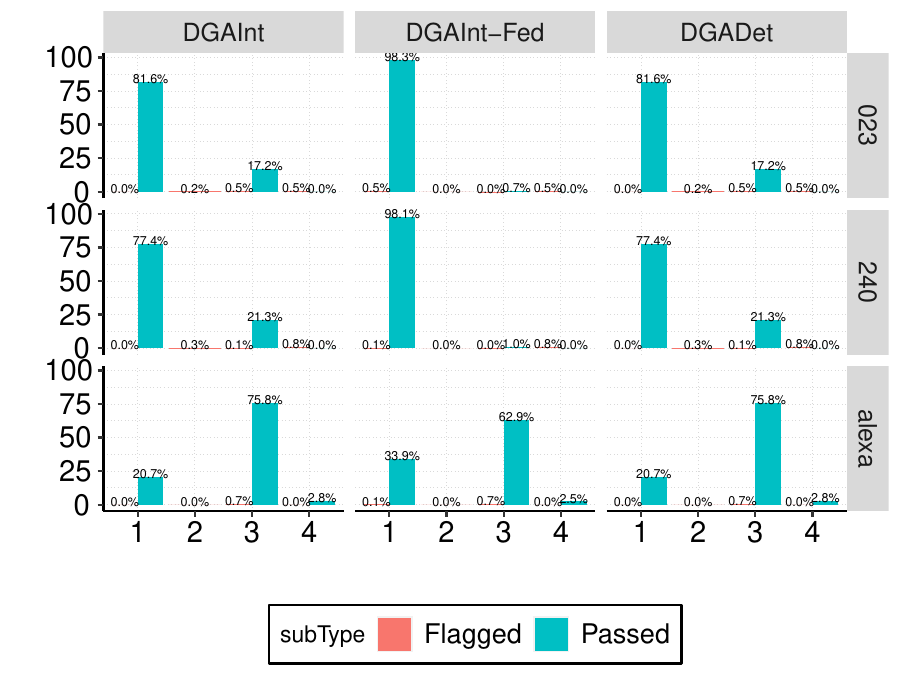}
    \caption{Benign}
    \label{fig:RatioPktBenign}
    \end{subfigure}%
    \begin{subfigure}{0.5\textwidth}
    \centering
    \includegraphics[scale=0.4]{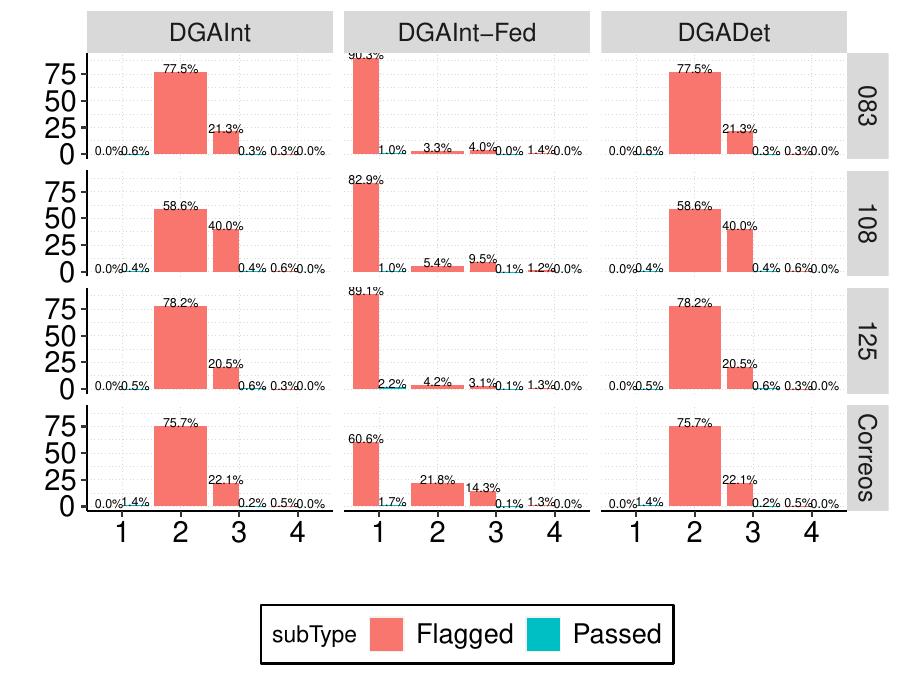}
    \caption{Infected}
    \label{fig:RatioPktInfected}
    \end{subfigure}
    \label{fig:RatioPkt}
    \caption{\ \\ Ratio of Packets Process per Test}
\end{figure}

Figures \ref{fig:RatioPktBenign} and \ref{fig:RatioPktInfected} show the percentage of packets by the phase they left the pipeline in the three DGA approaches, for the \textit{Benign} and \textit{Infected} files. In both cases, the use of the feedback loop in the \textit{DGAInt-Fed} leads to higher ratios of packets being processed at the blacklist phase - \textit{Phase 1}. Also, as expected the query analysis ratio in \textit{Phase 2} is able to identify a high percentage of packets (above 58\%). \textit{Phase 3} with the employment of DGA is also responsible for classifying a significant portion of packets depending if the file is benign or malign, above 70\% and above 20\%, respectively. 

\begin{figure}[!hbt]
    \begin{subfigure}{0.5\textwidth}
    \centering
    \includegraphics[scale=0.4]{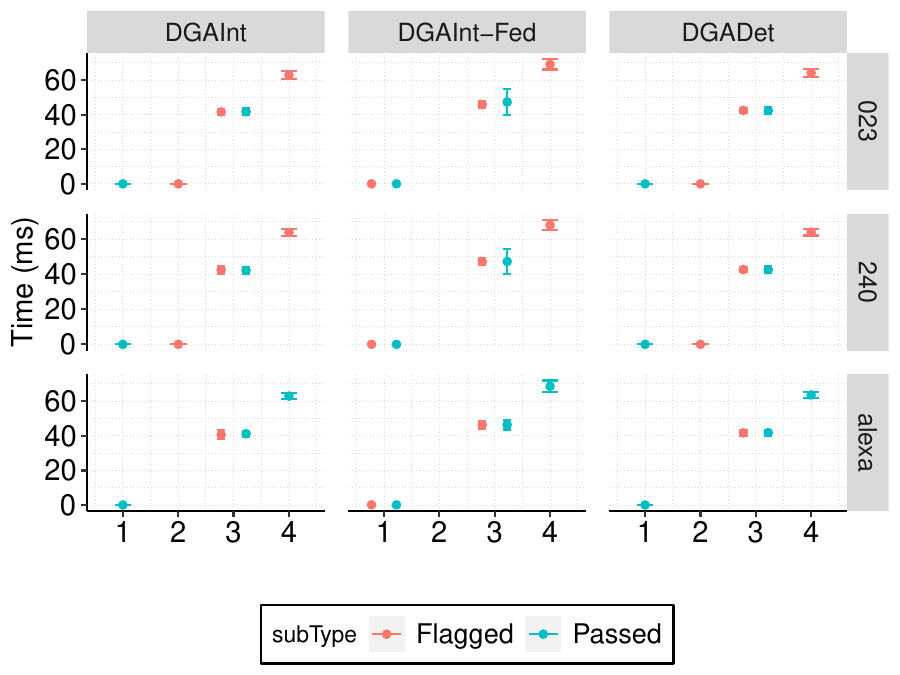}
    \caption{Benign}
    \label{fig:TimeBenign}
    \end{subfigure}%
    \begin{subfigure}{0.5\textwidth}
    \centering
    \includegraphics[scale=0.4]{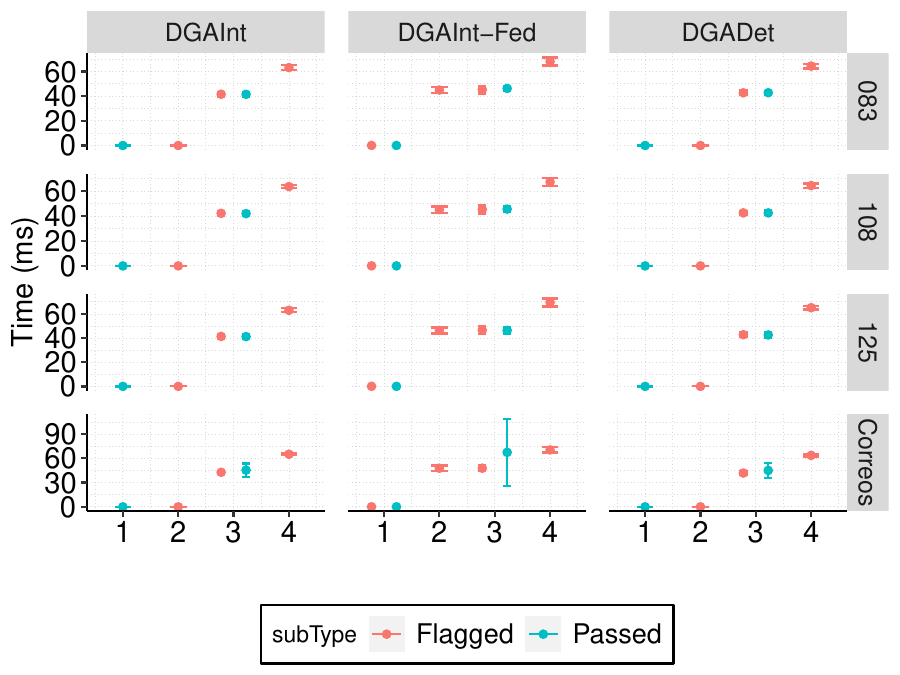}
    \caption{Infected}
    \label{fig:TimeInfected}
    \end{subfigure}
    \label{fig:Time}
    \caption{\ \\ Processing Time per Phase and Test }
\end{figure}
Figure~\ref{fig:TimeBenign} and \ref{fig:TimeInfected} illustrate the processing time of packets in each phase. The phase with higher processing time is the one associated with the \textit{RandomForest} ML model, as expected - \textit{Phase 4}. Indeed the difference is around 60ms between the performance of \textit{Phase 1} and \textit{Phase 4}, in the Benign and Infected files. In Figure~\ref{fig:TimeBenign} there are also some flagged packets that are classified as belonging to malicious requests. In the same line as the ML model, \textit{Phase 3} is also the one that takes more time due to the invocation of the DGA algorithms. In the Infected test case, as depicted in Figure~\ref{fig:TimeInfected}, the use of the feedback loop introduces additional processing, which impacts processing time in \textit{Phase 3} and \textit{Phase 4}. %
The lower times in \textit{Phase 1} and \textit{Phase 2} demonstrate the relevance of supporting these filtering functionalities towards highly efficient filtering approaches, with the possibility of being incorporated into existing security mechanisms like firewalls.

\subsection{MONDEO Pipeline Overhead}
\label{sec:evalResults:results:Overhead}
This subsection details the overhead results of the pipeline in terms of used resources: CPU, memory and the exchanged traffic in each approach.

\begin{figure}[!htbp]
  \centering
	\includegraphics[width=0.70\columnwidth]{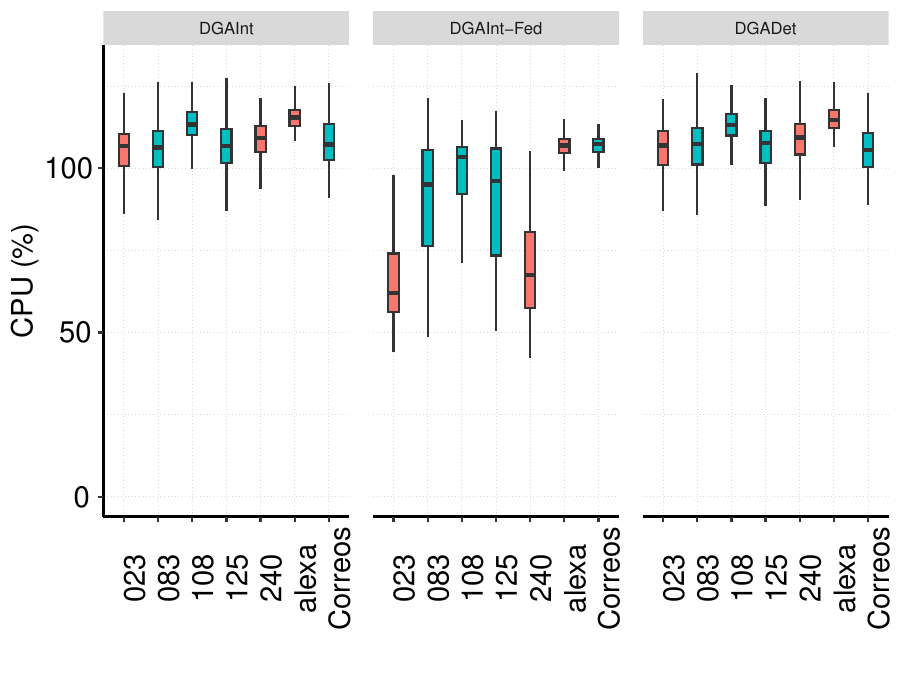}
	\caption{CPU usage ratio}
	\label{fig:CPUusage}
\end{figure}

The reported values for the CPU and memory utilisation are obtained from the docker statistics, each was collected in regular intervals (each 2s) during the experiments.

Figure~\ref{fig:CPUusage} depicts the CPU utilisation ratio in the diverse tests. The approaches, leading to higher CPU utilisation are the ones that do not use the feedback loop, namely the \textit{DGAInt} and the \textit{DGADet}. The reason is associated with a higher number of packets being processed in \textit{Phase 3} and \textit{Phase 4}, as previously discussed.

\begin{figure}[!htbp]
  \centering
	\includegraphics[width=0.70\columnwidth]{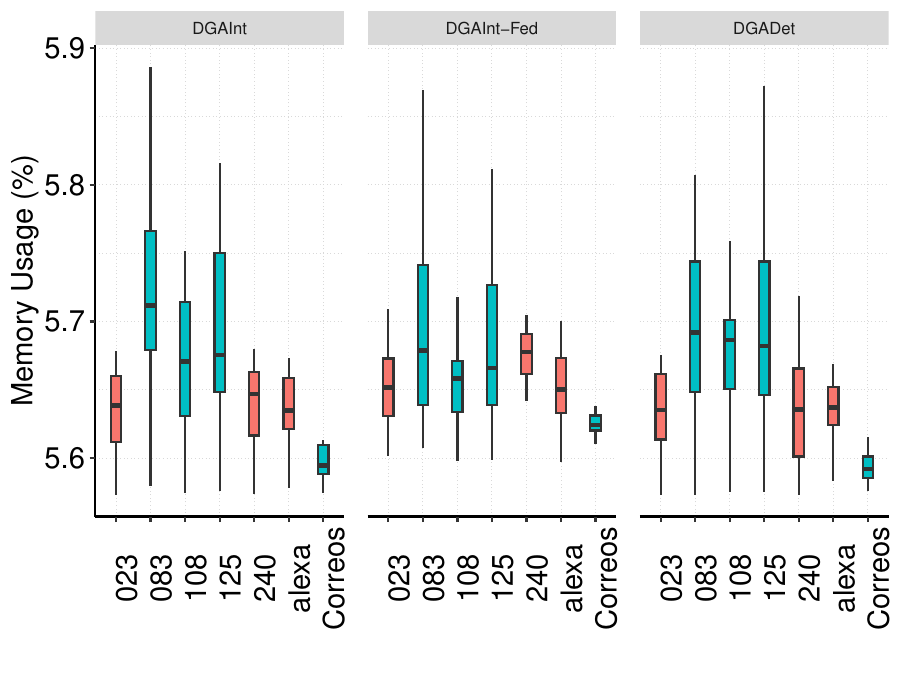}
	\caption{Memory usage ratio}
	\label{fig:MEMusage}
\end{figure}

As depicted in Figure~\ref{fig:MEMusage}, the memory usage ratio is low (below 6\%) in all the tests. The difference between the \textit{Benign} and \textit{Infected} files is neglectable (below 0.1\%), which demonstrates that MONDEO has a low memory footprint, despite having high ratios of CPU usage.

\begin{figure}[!htbp]
  \centering
	\includegraphics[width=0.70\columnwidth]{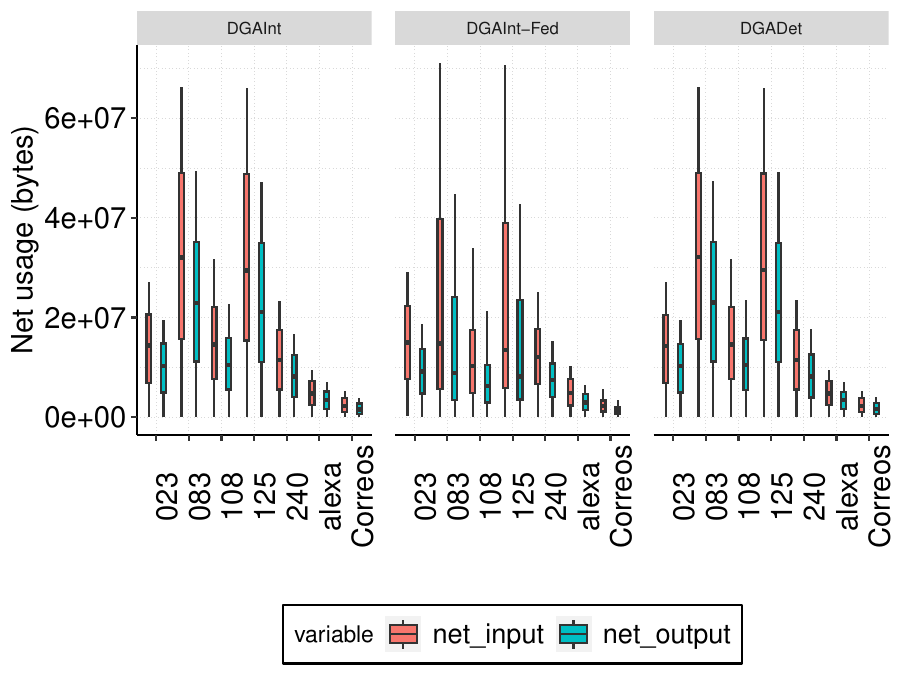}
	\caption{Input and output overhead in bytes}
	\label{fig:IOImpact}
\end{figure}

Figure~\ref{fig:IOImpact} depicts the impact regarding the exchanged traffic within each test. This metric translates the amount of information (packets) that is sent to the dockers where MONDEO is implemented and the respective answers. According to the evaluated scenario, there is a tendency to have less exchange of information with the feedback loop. This is due to the fact of avoiding communication with the docker running the DGA service, impacting, therefore the input and output traffic in both containers.

\section{Conclusions}
\label{sec:conclusions}

This paper proposes and validates \texttt{MONDEO} as a multistage approach for botnet detection, targeting malware relies on the DNS protocol and DGA obfuscation technique. \texttt{MONDEO} is a flexible and scalable mechanism that does not require software deployment or configuration in mobile devices. Thus, it is suitable to be implemented by network operators to combat such types of malware.

\texttt{MONDEO} relies on well-known detection approaches employed for botnet detection, such as DGA and flow information characteristics, to process streams of packets with high efficiency. \texttt{MONDEO} also supports white and blacklist that can use feedback loops to improve efficiency.

The evaluation results demonstrate the suitability of \texttt{MONDEO} to be deployed in network infrastructures of operators with minimal overhead and with accurate precision levels. 

Our next steps focus on the integration of policies and defence mechanisms, such that the results are mapped to mitigation tactics. Examples include blacklisting devices, and enabling dynamic policies through policy control. In addition, we will work with encrypted DNS traffic that is transmitted using HTTPS protocol.

%



\section*{Acknowledgments}
This work is funded by project AIDA (POCI-01-0247-FEDER045907), co-financed by the European Regional Development Fund (ERDF) through the Operational Program for Competitiveness and Internationalisation (COMPETE 2020) and by the Portuguese Foundation for Science and Technology (FCT) under CMU Portugal.
This work is funded by the FCT - Foundation for Science and Technology, I.P./MCTES through national funds (PIDDAC), within the scope of CISUC R\&D Unit - UIDB/00326/2020 or project code UIDP/00326/2020.


\bibliographystyle{IEEEtran}
\bibliography{MONDEO-Sec}

\end{document}